\documentclass[fleqn,twoside]{article}
\usepackage{espcrc2,epsfig, amssymb, longtable}

\usepackage{graphicx}

\newcommand{\AmS}{{\protect\the\textfont2
  A\kern-.1667em\lower.5ex\hbox{M}\kern-.125emS}}
\title{The Neutrinoless Double Beta Decay: The Case for Germanium Detectors \thanks{Invited contribution at the XXX
International Meeting on Fundamental Physics, IMFP2002, February
2002, Jaca, Spain.}}

\author{{A.~Morales\thanks{amorales@posta.unizar.es} and J.~Morales}
\\Laboratory of Nuclear Physics and
        High Energy Physics. Faculty of Science, University of Zaragoza, \\
        Pedro Cerbuna 12, 50009 Zaragoza, Spain}

\begin{document}

\begin{abstract}
\end{abstract}

\maketitle

\section{Motivation and Introduction}
In the Standard Model of Particle Physics neutrinos are strictly
massless, although there is no theoretical reason for such
prejudice. However, there exists conclusive experimental evidence
that the neutrino has a non-zero mass, as deduced from the
neutrino flavour oscillation observed in atmospheric and solar
neutrino experiments. On the other hand, galaxy formation requires
the presence (although small) of hot, non-baryonic dark matter
particles, like non-zero mass neutrinos to match the observed
spectral power.

In the Standard Model, neutrinos and antineutrinos are supposed to
be different particles, but no experimental proof has been
provided so far. The nuclear double beta decay addresses both
questions: whether the neutrino is self-conjugated and whether it
has a Majorana mass. The most direct way to determine if neutrinos
are Majorana particles is to explore, in potential nuclear double
beta emitters (A, Z), if they decay without emitting neutrinos
$\rightarrow$ (A, Z+2) + $2e^{-}$, violating the lepton number
conservation. For this non-standard $2\beta 0\nu$ process to
happen, the emitted neutrino in the first neutron decay must be
equal to its antineutrino and match the helicity of the neutrino
absorbed by the second neutron. Phenomenologically that implies
the presence of a mass term or a right-handed coupling. A
well-known argument of Schechter and Valle \cite{Sch1} shows that
in the context of any gauge theory, whatever mechanism be
responsible for the neutrinoless decay, a Majorana neutrino mass
is required. Moreover \cite{Kay1}, the observation of a $2\beta
0\nu$ decay implies a lower bound for the neutrino mass, i.e. at
least one neutrino eigenstate has a non-zero mass.

Another form of neutrinoless decay, (A, Z) $\rightarrow$ (A, Z+2)
+ $2e^{-}+\chi$ may reveal also the existence of the Majoron
($\chi$), the Goldstone boson emerging from the spontaneous
symmetry breaking of B--L, of most relevance in the generation of
Majorana neutrino masses and of far-reaching implications in
Astrophysics and Cosmology. Moreover, current neutrino physics
results have put on the front-line the double beta decay issue as
a probe to explore and elucidate important open questions left
unanswered by the oscillation experiments, namely the
determination of the absolute mass scale.

In fact, the recent results of neutrino oscillation experiments
indicate that next generation neutrinoless DBD experiments
(besides answering the long standing question of the Majorana
nature of neutrinos and the lepton number non-conservation) will
provide information on the type of mass spectrum, the absolute
neutrino mass scale and possibly the CP violation \cite{pascoli}.
The data from Super-K, SNO as well as from a large body of
previous neutrino oscillation experiments (Homestake, GALLEX,
SAGE, Kamiokande), together with their theoretical analysis,
clearly imply that neutrinos have indeed non-zero masses. However,
neutrino oscillation experiments determine only the mass squared
differences. Most of the models conclude that next-generation DBD
experiments with mass sensitivities of the order of 10 meV may
found the Majorana neutrino with a non-zero effective electron
neutrino mass, if the neutrino is selfconjugate and the neutrino
mass spectrum is of the quasi-degenerate type or it has inverted
hierarchy. Majorana massive neutrinos are common predictions in
most theoretical models, and the value of a few $10^{-2}$ eV
predicted for its effective mass, if reached experimentally -as
expected- will test its Majorana nature. DBD experiments with even
better sensitivities (of the order of meV) will be essential to
fix the absolute neutrino mass scale and possibly to provide
information on CP violation.

Neutrinoless DBD are extremely rare processes. Their experimental
investigation require a large amount of DBD emitter, in
low-background detectors with capability for selecting reliably
the signal from the background. Detectors should have a sharp
energy resolution, or good tracking of particles, or other
discriminating mechanisms. There are several natural and enriched
isotopes that have been used in experiments with tens of
kilograms. Some of them could be produced in amounts large enough
to be good candidates for next generation experiments. The choice
of the emitters should be made also according to its two-neutrino
half-life (which could limit the ultimate sensitivity of the
neutrinoless decay), according to its nuclear factor-of-merit and
according to the experimental sensitivity that the detector can
achieve in the case that the emitter is, at the same time, the
detector. For obvious reasons it is important to explore various
emitters and different techniques. Finally the extremely low
background required for high sensitivity double beta decay
searches requires to develop techniques for identifying, reducing
and suppressing the background of all types and origins in the
detectors and their environments . All these conditions set the
strategies to search for the neutrinoless double beta decay.

The expected signal rate depends on the nuclear matrix element but
the dispersion of results of the current calculations makes
uncertain the interpretation of the experimental output.
Improvements of the precision in theoretical evaluations of the
matrix elements are essential. Experimental studies of nuclear
structures, relevant to DBD, will help to perform adequate
calculations of the matrix elements. The exploration of the
conventional two-neutrino double beta decay of several potential
double beta emitters and its comparison with theory must serve to
help in determining the most suitable nuclear model. That
improvements will hopefully provide accurate nuclear matrix
elements for the neutrinoless DBD which are crucial for extracting
the effective Majorana mass parameter.

In this talk we will sketch the guidelines in the search for
double beta decay, in particular in the neutrinoless channel, and
illustrate the strategies by choosing the case of germanium
detectors, used in double beta searches as early as 1967. The
neutrino effective mass bound has been steadily decreasing along
the last decades, the stringent bound being provided by the
germanium experiments. Consequently we will put the emphasis in
the perspectives of the Ge$^{76}$ case for reaching the ten (and
less) millelectronvolt scale for the effective neutrino Majorana
mass in the next generation experiments.

The paper is organized as follows: Section 2 will resume the main
expressions for the half lives and mass bounds. Section 3 will
describe the experimental strategies. Section 4 will describe the
Ge experiments and give a comprehensive Table of the current
situation with other emitters. Also the bounds of the Majorana
effective neutrino mass will be shown. Finally Section 5 will
present the current prospects with Ge-emitters as an example to
illustrate the quest for the millielectronvolt mass sensitivity.

\section{The Double beta decay modes}

The two-neutrino decay mode (A, Z) $\rightarrow$ \\ (A, Z+2) +
$2e^{-}+2\overline{\nu}_{e}$ is a conventional \cite{Kay1},
although rare, second order weak process $(2\beta 2\nu)$, allowed
within the Standard Model. The half-lives are customary expressed
as $[T_{1/2}^{2\nu}\ (0^+\rightarrow 0^+ )^{-1}=G_{2\nu} \mid
M_{GT}^{2\nu} \mid^{2}$, where $G_{2\nu}$ is an integrated
kinematical factor \cite{Doi1} and $M_{GT}^{2\nu}$ the nuclear
double Gamow Teller matrix element.

The neutrinoless decay half-life (as far as the mass term
contribution is concerned) is expressed as
$(T_{1/2}^{0\nu})^{-1}=F_N\langle m_\nu\rangle^2/m_e^2$.
$F_N\equiv G_{0\nu} \mid M^{0\nu} \mid ^2$ is a nuclear
factor-of-merit and $M^{0\nu}$ is the neutrinoless nuclear
matrix-element, $M^{0\nu}=M_{GT}^{0\nu}-(g_V/g_A)^2\ M_F^{0\nu}$,
with $M_{GT,F}^{0\nu}$ the corresponding Gamow-Teller and Fermi
contributions. $G_{0\nu}$ is an integrated kinematic factor
\cite{Doi1}. The quantity $\langle m_\nu\rangle =\Sigma
\lambda_jm_jU_{ej}^2$ is the so-called effective neutrino mass
parameter, where $U_{ej}$ is a unitary matrix describing the
mixing of neutrino mass eigenstates to electron neutrinos,
$\lambda_j$ a CP phase factor, and $m_j$ the neutrino mass
eigenvalue.

Concerning the neutrino mass question, the discovery of a $2\beta
0\nu$ decay will tell that the Majorana neutrino has a mass equal
or larger than $\langle m_\nu\rangle
=m_e/(F_NT_{1/2}^{0\nu})^{1/2}$ eV, where $T_{1/2}^{0\nu}$ is the
neutrinoless half life. On the contrary, when only a lower limit
of the half-life is obtained (as it is the case up to now), one
gets only an upper bound on $\langle m_\nu\rangle$, but not an
upper bound on the masses of any neutrino. In fact, $\langle
m_\nu\rangle_{exp}$ can be much smaller than the actual neutrino
masses. The $\langle m_\nu\rangle$ bounds depend on the nuclear
model used to compute the $2\beta 0\nu$ matrix element. The
$2\beta 2\nu$ decay half-lives measured till now constitute
bench-tests to verify the reliability of the nuclear matrix
element calculations which, obviously, are of paramount importance
to derive the Majorana neutrino mass upper limit. As stated in the
Introduction the wide spread of the nuclear matrix elements
calculation for a given emitter is the main source of uncertainty
in the derivation of the neutrino mass bound. In the case of
$^{76}$Ge we quote the $2\beta0\nu$ nuclear merits
F$^{0\nu}_{N}(y^{-1})$, according to various nuclear models (see
Table \ref{nuclearmerits}).

\begin{table}[ht]
\caption{$2\beta0\nu$ nuclear merits F$^{0\nu}_{N}(y^{-1})$,
according to various nuclear models} \label{nuclearmerits}
\begin{tabular}{|ll|cc|}
\hline
Ref. & F$^{0\nu}_{N}(y^{-1}$) & Ref. & F$^{0\nu}_{N}(y^{-1}$) \\
\hline
 \cite{X1}  & 1.12 $\times 10^{ - 13}$ & \cite{X7a} & 7.33 $\times 10^{ - 14}$ \\
 \cite{X2} & 1.12 $\times 10^{ - 13}$  &  \cite{X7b} & 1.42 $\times 10^{ - 14}$\\
 \cite{X3} & 1.87 $\times 10^{ - 14}$  & \cite{X8} & 5.8 $\times 10^{ - 13}$\\
 \cite{X4} & 1.54 $\times 10^{ - 13}$ & \cite{X9} & 1.5 $\times 10^{ - 14}$\\
 \cite{X5} & 1.13 $\times 10^{ - 13}$ & \cite{X10} &  9.5 $\times 10^{ - 14}$ \\
 \cite{X6} & 1.21 $\times 10^{ - 13}$ &  & \\
\hline
\end{tabular}
\end{table}

\section{Searching for Double Beta Decays: Guidelines}

The experimental signatures of the nuclear double beta decays are
in principle very clear: In the case of the neutrinoless decay,
one should expect a peak (at the Q$_{2\beta}$ value) in the
two-electron summed energy spectrum, whereas two continuous
spectra (each one of well-defined shape) will feature the
two-neutrino and the Majoron-neutrinoless decay modes (the first
having a maximum at about one third of the Q value, and the latter
shifted towards higher energies). In spite of such characteristic
imprints, the rarity of the processes under consideration make
very difficult their identification. In fact, double beta decays
are very rare phenomena, with two-neutrino half-lives as large as
$10^{18}$ y to $10^{24}$ y and with neutrinoless half-lives as
long as $10^{25}$ y (and beyond), as the best lower limit stands
by now. Such remotely probable signals have to be disentangled
from a (much bigger) background due to natural radioactive decay
chains, cosmogenic-induced activity, and man-made radioactivity,
which deposit energy on the same region where the $2\beta$ decays
do it but at a faster rate. Consequently, the main task in
$2\beta$-decay searches is to diminish the background by using the
state-of-the-art ultralow background techniques and, hopefully,
identifying the signal.

To measure double beta decays three general approachs have been
followed: geochemical, radiochemical and direct counting
measurements. In the geochemical experiments, isotopic anomalies
in noble gases daughter of $2\beta$ decaying nucleus over
geological time scales are looked for. Determination of the gas
retention age of the ore is important. They are inclusive
$2\nu+0\nu$ measurements, not distinguishing $2\nu$ from $0\nu$
modes.

However, when $T_{1/2,exp.meas.}^{2\nu+0\nu(geoch.)} \ll
T_{1/2,exp.bound}^{0\nu(direct)}$, most of the decay is through
$2\nu$ mode. The finite half-lives measured geochemically in the
cases of $^{82}$Se, $^{96}$Zr, $^{128,130}$Te, $^{238}$U can be
regarded as $2\beta2\nu$ half-life values. Also the $T_{1/2}$
values measured geochemically can be taken as a bound for
$T_{1/2}^{0\nu}$, because $T_{1/2}^{0\nu}$ (or the half-life of
whatever decay mode) cannot be shorter than $T_{1/2,exp.}$.
Consequently, bounds on $\langle m_\nu\rangle$ can be derived from
geochemical half-life measurements.

Another way to look for double beta decays are the radiochemical
experiments, by noticing that when the daughter nuclei of a double
beta emitter are themselves radioactive, they can be accumulated,
extracted and counted. If the daughter has a half-life much
smaller than $10^{9}$y and has no other long-lived parents, its
presence can be only due to $2\beta$. Parent minerals must have
been isolated before 1940. Noticeable examples are that of
$^{238}U \to \, ^{239}Pu$ (88 y, $\alpha$ decay) and $^{244}Pu \to
\, ^{244}Cm$ (18 y, $\alpha$ decay).

Most of the recent activity, however, refers to direct counting experiments,
which measure the energy of the $2\beta$ emitted electrons and so the spectral
shapes of the $2\nu$, $0\nu$, and $0\nu \chi$ modes of double beta decay.
Some experimental devices track also the electrons (and other charged particles),
measuring the energy, angular distribution, and topology of events. The tracking
capabilities are essential to discriminate the $2\beta$ signal from the background.
The types of detectors currently used are:

\begin{itemize}
\item Calorimeters where the detector is also the $2\beta$ source
(Ge diodes, scintillators --CaF$_2$, CdWO$_4$--, thermal
detectors, ionization chambers). They are calorimeters which
measure the two-electron sum energy and discriminate partially
signal from background by pulse shape analysis (PSD). Notable
examples of calorimeters are IGEX, H/M and MIBETA.
 \item
Tracking detectors of source$\neq$detector type (Time Projection
Chambers TPC, drift chambers, electronic detectors). In this case,
the $2\beta$ source plane(s) is placed within the detector
tracking volume, defining two --or more-- detector sectors.
Leading examples of tracking devices are the NEMO series and
ELEGANTS
 \item
Tracking calorimeters: They are tracking devices where the
tracking volume is also the $2\beta$ source, for example a Xenon
TPC.
\end{itemize}

Well-known examples of $2\beta$ emitters measured in direct counting
experiments are $^{48}$Ca, $^{76}$Ge, $^{96}$Zr, $^{82}$Se, $^{100}$Mo,
$^{116}$Cd, $^{130}$Te, $^{136}$Xe, $^{150}$Nd.

The strategies followed in the $2\beta$ searches are varied.
Calorimeters of good energy resolution and almost 100\% efficiency
(Ge-detectors, Bolometers) are well suited for $0\nu$ searches.
They lack, obviously, the tracking capabilities to identify the
background on an event-by-event basis but they have, in favour,
that their sharp energy resolution do not allow the leakage of too
many counts from ordinary double beta decay into the neutrinoless
region, and so the sensitivity intrinsic limitation is not too
severe. The identification capabilities of the various types of
chambers make them very well suited for $2\nu$ and $0\nu \chi$
searches. However, their energy resolution is rather modest and
the efficiency is only of a few percent. Furthermore, the ultimate
major background source in these devices when looking for $2\beta
0\nu$ decay will be that due to the standard $2\beta 2\nu$ decay.
The rejection of background provided by the tracking compensates,
however, the figure of merit in $0\nu$ searches.

Modular calorimeters can have reasonable amounts of $2\beta$
emitters (Heidelberg/Moscow, IGEX, MIBETA and CUORICINO
experiments) or large quantities (like CUORE and Majorana).
Tracking detectors, instead, cannot accommodate large amounts of
$2\beta$ emitters in the source plate. Recent versions of tracking
devices have 10 kg and more (NEMO3).

The general strategy followed to perform a neutrinoless double
beta decay experiment is simply dictated by the expression of the
half-life (T$_{1/2}^{0\nu}\simeq ln2\times\frac{N.t}{S}$) where N
is the number of 2$\beta$ emmiter nuclei and S the number of
recorded counts during time t (or the upper limit of double beta
counts consistent with the observed background). In the case of
taking for S the background fluctuation one has the so-called
detector factor-of-merit or neutrinoless sensitivity which for
source$=$detector devices reads $F_D=4.17 \times
10^{26}(f/A)(Mt/B\Gamma)^{1/2} \varepsilon_{\Gamma}$ years where B
is the background rate (c/keV kg y), M the mass of $2\beta$
emitter (kg), $\varepsilon_{\Gamma}$ the detector efficiency in
the energy bin $\Gamma$ around Q$_{2\beta}$ ($\Gamma =$ FWHM) and
t the running time measurement in years (f is the isotopic
abundance and A the mass number). The other guideline of the
experimental strategy is to choose a $2\beta$ emitter of large
nuclear factor of merit $F_N=G_{0\nu} \mid M^{0\nu} \mid^2$, where
the kinematical factor qualifies the goodness of the $\rm
Q_{2\beta}$ value and $M^{0\nu}$ the likeliness of the transition.
Notice that the upper limit on $\langle m_\nu \rangle$ is given by
$\langle m_\nu\rangle\le\langle m_e\rangle/(F_D F_N)^{1/2}$, or in
terms of its experimental and theoretical components \\ $\langle
m_\nu \rangle\le2.5\times
10^{-8}\times(A/f)^{1/2}\times(B\Gamma/Mt)^{1/4}\times\epsilon_{\Gamma}^{-1/2}\times
G_{0\nu}^{-1/2}\times\mid M^{0\nu}\mid^{-1} eV$

\section{Experimental Searches: Overview of Recent Results. The Germanium Case}

The current status of the double beta decay searches is sketched
in Table \ref{tablagorda} where the main features, parameters and
results of the experiments are summarized. From the list of
emmiters and experiments quoted in Table \ref{tablagorda} the
germanium case has been chosen to illustrate in some detail a
typical strategy in double beta searches.

\begin{table*}[t]
\caption{Theoretical half-lives $T_{1/2}^{2\nu}$ in some
representative nuclear models.} \label{teorhf} \footnotesize
\begin{tabular*}{\textwidth}{l@{\extracolsep{\fill}}llllllll}
\hline
                 & \multicolumn{8}{c|}{\sl{Theory}}
                 \\ \hline

                 & \multicolumn{3}{|c|}{SM}
                 & \multicolumn{2}{c|}{QRPA}
                 & \multicolumn{1}{c|}{$1^+D$}
                 & \multicolumn{1}{c|}{OEM}
                 & \multicolumn{1}{c|}{MCM}
                 \\  \hline

                 & \cite{X4}
                 & \cite{Hax2}
                 & \cite{X9}
                 & \cite{X3}
                 & \cite{X1}
                 & \cite{Aba1}
                 & \cite{X10}
                 & \cite{Suh2}
                 \\  \hline
\\
$^{48}$Ca($10^{19}$y) & 2.9 & 7.2 & 3.7 &  &  &  &  &
\\
 $^{76}$Ge($10^{21}$y) & 0.42 & 1.16 & 2.2 & 1.3 & 3.0 & &
0.28 &1.9
\\
 $^{82}$Se($10^{20}$y) & 0.26 & 0.84 & 0.5 & 1.2 & 1.1 & 2.0
& 0.88 & 1.1

\\
 $^{96}$Zr($10^{19}$y) &  &  &  & 0.85 & 1.1 &  &  & .14-.96
\\
 $^{100}$Mo($10^{19}$y) &  &  &  & 0.6 & 0.11 & 1.05 & 3.4 &
0.72
\\
 $^{116}$Cd($10^{19}$y) &  &  &  &  & 6.3 & 0.52 &  & 0.76
\\
 $^{128}$Te($10^{24}$y) & 0.09 & 0.25 &  & 55 & 2.6 & 1.4 &  &
\\
 $^{130}$Te($10^{21}$y) & 0.017 & 0.051 & 2.0 & 0.22 & 1.8 & 2.4 & 0.1 &
\\
 $^{136}$Xe($10^{21}$y) &  &  & 2.0 & 0.85 & 4.6 &  &  &
\\
 $^{150}$Nd($10^{19}$y) &  &  &  &  & 0.74 &  &  &
\\ \hline \normalsize
\end{tabular*}
\end{table*}

There exist two experiments in operation looking for the double
beta decay of $^{76}$Ge. They both employ several kilograms of
enriched $^{76}$Ge (86\%) in sets of detectors: the IGEX
Collaboration experiment \cite{X11} (a set of three detectors of
total mass 6.3 kg) in the Canfranc Underground Laboratory (Spain)
and the Heidelberg/Moscow Collaboration experiment \cite{X12} (a
set of five detectors amounting to 10.2 kg) running in Gran Sasso.
Both experiments where designed to get the highest possible
sensitivity to look for the neutrinoless double beta decay of the
$^{76}$Ge: a large amount of the Ge-76 isotope, in detectors of
good energy resolution and very low radioactive background.

In the case of IGEX, the quest for an ultralow background started
with a thorough radiopurity screening of the materials to be used
in the detectors and in the inner components of the shielding. A
thick passive shield (of about roughly 50 cm of lead) was
employed. A continuous flux of clean nitrogen gas was injected
into the shield to evacuate the radon. On the other hand an active
veto shield rejected muon induced events and a neutron shield
completed the barrier against external sources of background. This
shielding will be described later on. A final step in the
reduction was obtained through the event selection via Pulse Shape
Discrimination (PSD) explained later on.

The FWHM energy resolutions of the three 2 kg IGEX detectors at
1333-keV are 2.16, 2.37, and 2.13 keV, and the energy resolution
of the summed data integrated over the time of the experiment was
$\sim4$keV at Q$=$2039 keV. The detectors cryostats were made in
electroformed copper. The cooper part of the cryostat were
produced by special techniques to eliminate Th and Ra impurities.
All other components were made from radiopure materials.  The
first stage FET (mounted on a Teflon block a few centimeters apart
from the inner contact of the crystal) was shielded by 2.6 cm of
500 y old lead to reduce the background. Also the protective cover
of the FET and the glass shell of the feedback resistor were
removed for such purpose. Further stages of amplification were
located 70 cm away from the crystal. All the detectors have
preamplifiers modified for pulse shape analysis (PSD) for
background identification. Finally, special care was put in
maintain a long-term stability in gain, resolution, noise, count
rate, PSD signal, etc.

The data acquisition system had an independent spectroscopy chain
for each Ge detector. The threshold was set to 1.5 MeV in plastic
scintillators to register low energy deposits of muons and to 100
keV in Ge detectors to avoid unnecessary high count rates. For
each Ge event, the time elapsed since run started ($100\mu s$
ticks), the time elapsed since last veto signal ($20\mu s$ ticks),
the ADC channel numbers (range 0-8191, $\sim1$ keV/ch) and the
scope trace with pulse shape (500 points, 2ns/point) were
recorded. As far as stability on gain and resolution the energy
shifts at $\sim1.3$ MeV are smaller than 0.5 keV (typically 0.3
keV) over 2 months and the energy resolution at $\sim 1.3$ MeV was
2.2 - 2.5 keV over 2 months. The integrated resolution over $\sim
900$ days of data spread to $\sim 3.5$ keV.

The IGEX setup has been running in the Canfranc Underground
Laboratory (Spain) at 2450 m.w.e. Special attention deserves the
heavy shielding which was developed for this experiment. It
consist of a large shielding enclosing tightly the set of
detectors. First there is an innermost shield of 2.5 tons ($\sim
60$ cm cube) of archaeological lead (2000 yr old) --having a
$^{210}$Pb($^{210}$Bi) content of $<0.01$ Bq/kg--, where the 3
large detectors are fitted into precision-machined holes to
minimize the empty space around the detectors available to radon.
Nitrogen gas evaporated from liquid nitrogen, is forced into the
remaining free space to minimize radon intrusion. Surrounding the
archaeological lead block there is a 20-cm thick layer of low
activity lead ($\sim 10$ tons), sealed with plastic and cadmium
sheets. A cosmic muon veto and a neutron shield (20 cm thick made
from polyethylene bricks enlarged later on up to 40 cm plus 20 cm
of borated water, placed externally) close the assembly.

\begin{figure}[h]
\centerline{\includegraphics[width=7cm]{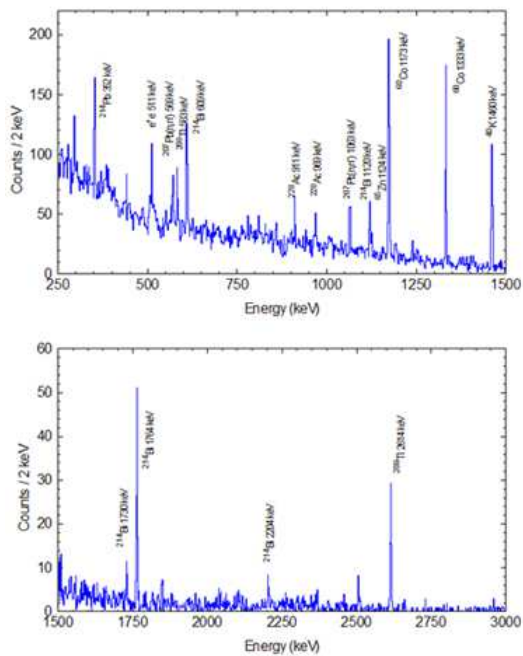}} \caption{Energy
spectrum of one IGEX detector (RG-II) for exposure of 2.75 kg y.}
\label{fig1}
\end{figure}

The energy spectrum of one IGEX detector is shown in the Figure
\ref{fig1}. Most of the background in the relevant Q$_{2\beta}$
region of 2039 keV is accounted for by cosmogenic activated nuclei
($^{68}$Ge and $^{60}$Co). The background recorded in the energy
region between 2.0 and 2.5 MeV is about 0.2 c/keV kg y prior to
PSD. Background reduction through Pulse Shape Discrimination
successfully eliminate multisite events, characteristic of
non-$2\beta$ events.

The rationale for PSD is quite simple: in large intrinsic Ge
detectors, the electric field increases by a factor or more than
10 from the inner conductor to the outer conductor, which are
almost 4 cm apart. Electrons and holes take 300~-~500 ns to reach
their respective conductors. The current pulse contributions from
electron and holes are displacement currents, and therefore
dependent on their velocities and radial positions. Accordingly,
events occurring at a single site ($\beta\beta$-decay events for
example) have associated current pulse characteristics that
reflect the position in the crystal where the event occurred. More
importantly, these single-site event frequently have pulse shapes
that differ significantly from those due to the most dominant
background events that produce electron-hole pairs at several
sites by multi-Compton-scattering process, for example.
Consequently, pulse-shape analysis can be used to distinguish
between these two types of energy depositions.

To develop PSD techniques it is helpful to work with a signal as
close as possible to the displacement current of the detector.
This allows the development of algorithms that do not depend
strongly on the preamplifier electronics in use. To this end, the
transfer function of the preamplifier and associated front-end
stage has been measured for each detector. This allows the
reconstruction of the displacement current and easy comparison to
computed pulse shapes.

Double-beta decay events will deposit energy at a single site in a
detector. Most background events will deposit energy at several
sites. Our models of the structure of the current pulse reveal
that single-site events will exhibit only one or two features, or
``lobes'', in more than 97\% of the cases. Multiple site events
will most often exhibit more than two lobes.

One PSD technique is to reject pulses having more than two
significant lobes or peaks. To detect lobes, a ``Mexican-hat''
filter of the proper width is applied to the pulse. This robust
method is nearly model-independent.  Some multiple-site events may
show only one or two lobes and will not be rejected by this
technique.  Use of this PSD method results in the rejection of
60\%--80\% of the IGEX background in the energy interval
2.0--2.5~MeV, down to less than $\sim0.07$c/keV.kg.y.

\begin{figure}[h]
\centerline{\includegraphics[width=7cm]{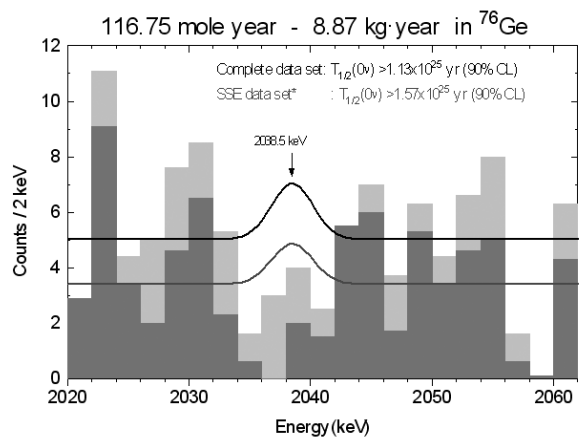}}
\caption{Two-electron summed energy spectrum of the IGEX
experiment around $\rm Q_{2\beta}=2039$ keV region} \label{fig2}
\end{figure}

The IGEX data histrograms for the $0\nu \beta \beta$ region are
shown in Fig. \ref{fig2}. The combined energy resolution is 4 keV.
PSD has only been applied to a fraction of data and so two
analysis were made, with and without inclusion of PSD. In the Fig.
\ref{fig2}, the light-gray area represents the complete data set
corresponding to 116.75 fiducial mole-years (8.87 kg.y) and the
dark-grey area represents the complete data set after the
application of PSD (only 4.64 kg.y of data out of the total 8.87
kg.y have been Pulse Shape analyzed --2.74 kg.y of detector RG2
data, with rejection factor of 60.44\% and 1.90 kg.y of detector
RG3 data with a rejection factor of 76.54\%--).

\begin{table}[ht]
\caption{IGEX Data bins for 8.87 kg.y in $^{76}$Ge}
\label{databins} \footnotesize
\begin{tabular}{ccc}
\hline E(low) keV & SSE data set & Complete data set\\ \hline
2020 & 2.9 & 2.9 \\
2022 & 9.1 & 11.1 \\
2024 & 3.4 & 4.4\\
2026 & 2.0 & 5.0 \\
2028 & 4.6 & 7.6 \\
2030 & 6.5 & 8.5 \\
2032 & 2.3 & 5.3 \\
2034 & 0.6 & 1.6 \\
2036 & 0.0 & 3.0 \\
2038 & 2.0 & 4.0 \\
2040 & 1.5 & 2.5 \\
2042 & 5.5 & 5.5 \\
2044 & 6.0 & 7.0 \\
2046 & 1.7 & 3.7 \\
2048 & 5.3 & 6.3 \\
2050 & 3.4 & 4.4 \\
2052 & 4.6 & 6.6 \\
2054 & 5.0 & 8.0 \\
2056 & 0.6 & 1.6 \\
2058 & 0.1 & 0.1 \\
2060 & 4.3 & 6.3 \\
\hline
Expected counts&13.6&20.08\\
Observed counts&4.1&11.1\\
Upper limit A &3.1 &4.3 \\
(90\%CL) &&\\\hline
ln2.Nt/A&$1.57\times10^{25}$y&$1.13\times10^{25}$y\\
 \hline
\normalsize
\end{tabular}
\end{table}

Table \ref{databins} shows the IGEX data for the mentioned
exposure of 8.87 kg.y in $^{76}Ge$ (expressed as number of counts
per 2 keV bin) in the region between 2020 and 2060 keV, for the
two sets of 8.87 kg.y of data with (first column) and without
(second column) application of PSD. Using the statistical
estimator recommended by the Particle Data Group, 90\% C.L.
half-life lower bounds of $T_{1/2}^{0\nu }\geq 1.13\times 10^{25}
y$ for the complete data set (dashed line) and of $T_{1/2}^{0\nu
}\geq 1.57\times 10^{25} y$ for the complete data set with PSD
(solid line) are obtained (see Table \ref{databins}) \cite{X11}.
Accordingly, the limit on the neutrino mass parameter is
0.38--1.55~eV for one data set and 0.33--1.31~eV for the other
data set as shown in Table \ref{pequena}. The uncertainties
originate from the spread in the calculated nuclear structure
parameters.

\begin{table}[ht]
\caption{$\langle m_{\nu}\rangle$ bounds resulting from IGEX
($T_{1/2}^{0\nu }\geq 1.57\times 10^{25} y$) according to various
representative nuclear models.} \label{pequena} \footnotesize
\begin{tabular}{llcc}
\hline & &\multicolumn{2}{c}{$\langle m_{\nu}\rangle$(eV)}\\
 & & Complete&SSE\\
F$_{N}$(y$^{-1}$) & Model &Data set&Data set\\ \hline
$1.56\times10^{-13}$ & WCSM\cite{X4}& 0.38 & 0.33 \\
$9.67\times10^{-15}$ & QRPA\cite{X3} & 1.55 & 1.31 \\
$1.21\times10^{-13}$ & QRPA\cite{X6} & 0.44 & 0.37 \\
$1.12\times10^{-13}$ & QRPA\cite{X1} & 0.45 & 0.39 \\
$1.41\times10^{-14}$ & SM\cite{X9} & 1.28 & 1.09 \\
 \hline
\normalsize
\end{tabular}
\end{table}



\begin{figure}[h]
\centerline{\includegraphics[width=7cm]{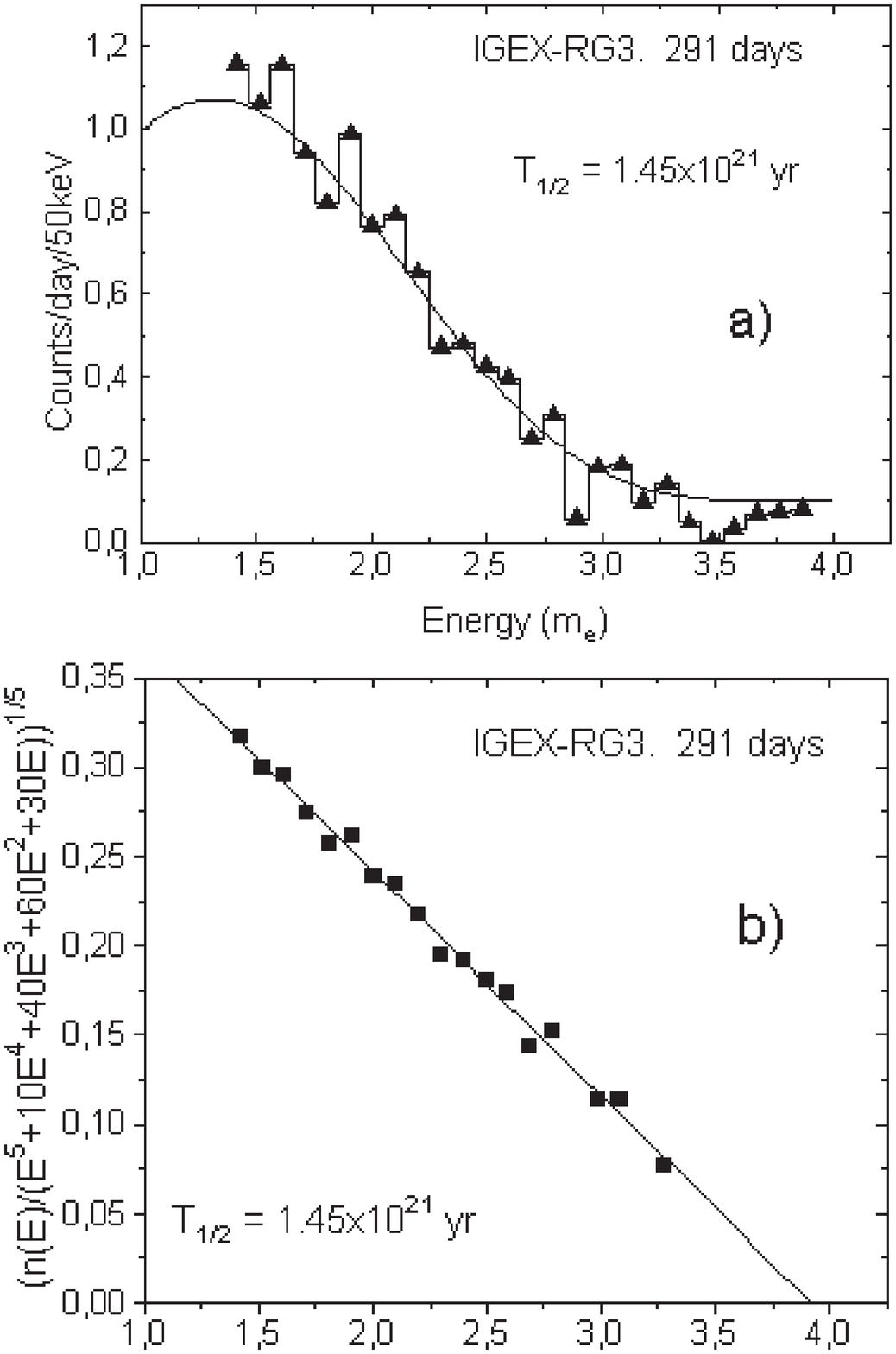}} \caption{a)
Best fit to the stripped data corresponding to a half-life
$T^{2\nu}_{1/2}=(1.45 \pm 0.20) \times 10^{21}$ y. b) Experimental
points fit the double beta Kurie plot} \label{fig3}
\end{figure}

Data from one of the IGEX detectors, RG--3 --which went
underground in Canfranc several years ago-- corresponding to 291
days, were used to set a value for the $2\nu$-decay mode half-life
by simply subtracting the MC-simulated background. Figure
\ref{fig3}-a shows the best fit to the stripped data corresponding
to a half-life $T^{2\nu}_{1/2}=(1.45 \pm 0.20) \times 10^{21}$ y,
whereas Figure \ref{fig3}-b shows how the experimental points fit
the double beta Kurie plot (see Ref. \cite{amnu98}).

\begin{table}[ht]
\caption{Limits on Neutrinoless Decay Modes} \label{limits}
\footnotesize
\begin{tabular}{llcc}
\hline \sl{Emitter} & \sl{Experiment} & $\sl T_{1/2}^{0\nu} >$ &
\sl{C.L.\%} \\ \hline
$^{48}$Ca & HEP Beijing & $1.1 \times 10^{22}$ y & 68 \\
$^{76}$Ge & MPIH/KIAE & $1.9 \times 10^{25}$ y & 90 \\
          & IGEX  & $1.6 \times 10^{25}$ y & 90 \\
$^{82}$Se & UCI & $2.7 \times 10^{22}$ y & 68 \\
          & NEMO 2 & $9.5 \times 10^{21}$ y & 90 \\
$^{96}$Zr & NEMO 2 & $1.3 \times 10^{21}$ y & 90 \\
$^{100}$Mo & LBL/MHC/UNM & $2.2 \times 10^{22}$ y & 68 \\
           & UCI & $2.6 \times 10^{21}$ y & 90 \\
           & Osaka & $6.5 \times 10^{22}$ y & 90 \\
           & NEMO 2 & $6.4 \times 10^{21}$ y & 90 \\
$^{116}$Cd & Kiev & $7 \times 10^{22}$ y & 90 \\
           & Osaka & $2.9 \times 10^{21}$ y & 90 \\
           & NEMO 2 & $5 \times 10^{21}$ y & 90 \\
$^{130}$Te & Milano & $2.1 \times 10^{23}$ y & 90 \\ $^{136}$Xe &
Caltech/UN/PSI & $4.4 \times 10^{23}$ y & 90 \\
 $^{136}$Xe & Rome
& $7 \times 10^{23}$ y & 90 \\
$^{150}$Nd & UCI & $1.2 \times
10^{21}$ y & 90 \\ \hline \normalsize
\end{tabular}
\end{table}

The Heidelberg/Moscow experiment operates five p-type HPGe
detector of enriched $^{76}$Ge (86\%) with a total active mass 0f
10.96 kg, corresponding to 125.5 mol of $^{76}$Ge. A detailed
description of the experiment is given in Ref. \cite{X12} and
references therein.

\begin{figure}[h]
\centerline{\includegraphics[width=7cm]{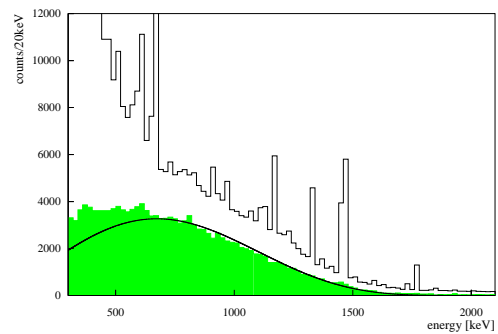}}
\caption{Energy spectrum of the five H/M detectors for 47.7 kg y
of exposure.} \label{fig4}
\end{figure}

\begin{figure}[h]
\centerline{\includegraphics[width=7cm]{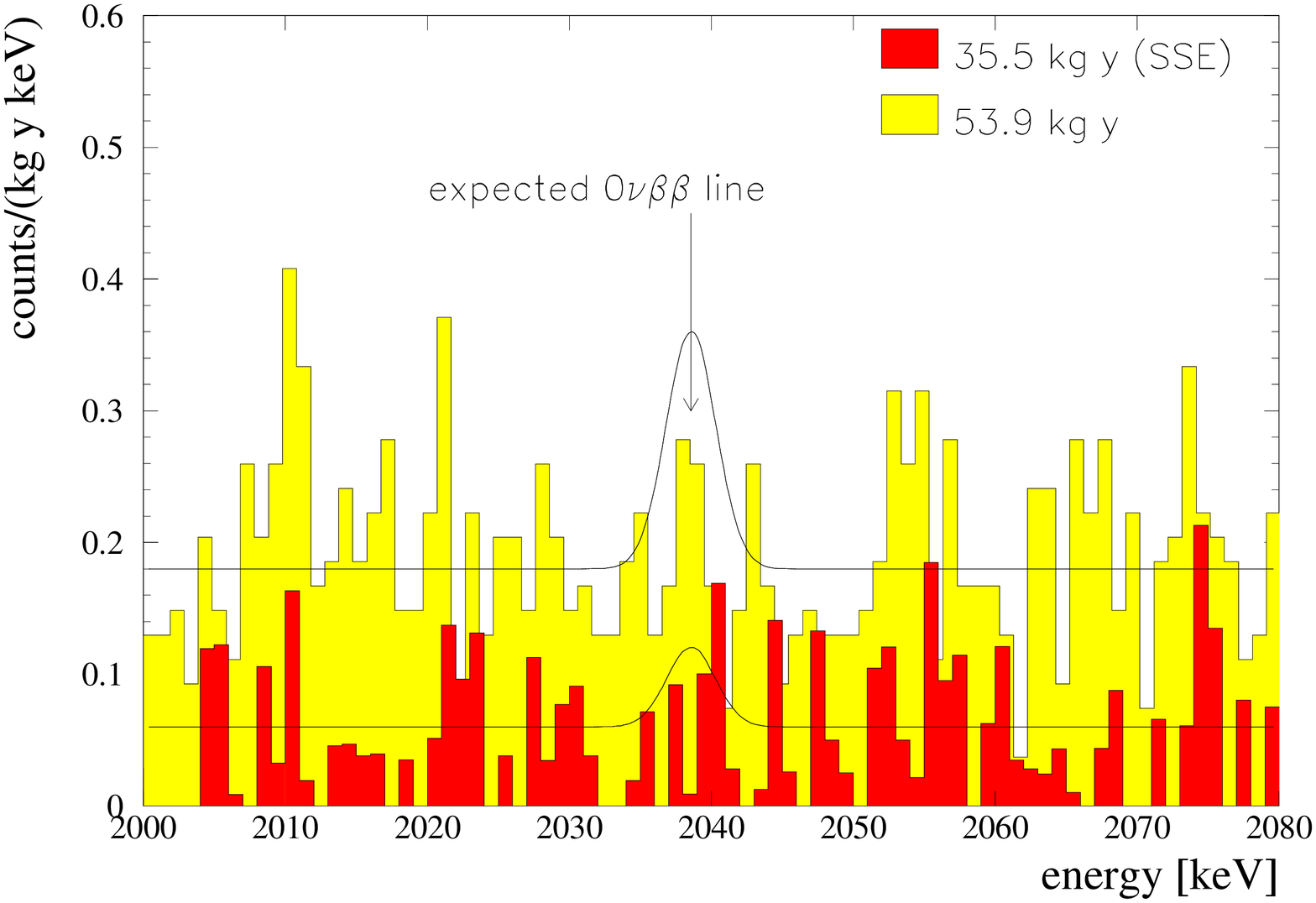}}
\caption{Two-electron summed energy spectrum of the H/M experiment
around Q$_{2\beta}$=2039 keV region.} \label{fig5}
\end{figure}
\begin{table*}[ht]
\caption{Neutrinoless half-lives in various Theoretical Models
(for the $\langle m_\nu \rangle$ term) $T_{1/2}^{0\nu}\langle
m_\nu \rangle^2$ values are given in $10^{24}$ (eV)$^2$ y.}
\label{models} \footnotesize
\begin{tabular*}{\textwidth}{l@{\extracolsep{\fill}}rrrrrrrrr}
\hline
  & $^{76}$Ge  & $^{82}$Se & $^{100}$Mo & $^{128}$Te & $^{130}$Te
  & $^{136}$Xe & $^{150}$Nd & $^{116}$Cd & $^{48}$Ca \\ \hline
  Weak Coupl. SM \cite{X4} & 1.67 & 0.58 &    &  4.01 & 0.16 &  &  &  &  \\
  $g_A=1.25 (g_A=1)$ & (3.3) & (1.2) &  & (7.8) & (0.31) &  &  &  &   \\
  Large Space SM \cite{X9} & 17.5 & 2.39 & & & & 12.1 & & & 6.25 \\
  QRPA \cite{X3} & 14 & 5.6 & 1.9 & 15 & 0.66 & 3.3 &  &  &  \\
  QRPA \cite{X2} & 2.3 & 0.6 & 1.3 & 7.8 & 0.49 & 2.2 & 0.034 & 0.49 &  \\
  QRPA \cite{X6} & 2.15 & 0.6 & 0.255 & 12.7 & 0.52 & 1.51 & 0.045 &  &  \\
  OEM \cite{X10} & 2.75 & 0.704 &  & 12.6 & 0.723 & 4.29 & 0.056 & 0.583  & \\
  QRPA with \cite{X7b} & 18.4 & 2.8 & 350 & 150 & 2.1 & 2.8 &  & 4.8 & 28 \\
  (without) np pair. \cite{X7a} & (3.6) & (1.5) & (3.9) & (19.2) & (0.86) &  & (4.7)& (2.4) \\
  \hline
\normalsize
\end{tabular*}
\end{table*}

Figure \ref{fig4} shows the recorded background spectrum of the
detectors of the H/M experiment corresponding to an exposure of
47.4 kg.y compared with the MonteCarlo simulation of the
background. After substraction, the resulting half-life for the
$2\nu\beta\beta$-decay at 68\% C.L. is
\begin{center}
$T^{2\nu}_{1/2}=\{1.55\pm0.01(stat)^{+0.19}_{-0.15}(syst)\}\times10^{21}$y
\end{center}
in agreement with the 1998 IGEX result \cite{amnu98}.

To derive the neutrinoless decay half-life limit from the H/M
experiment, the raw data of all five detectors as well as data
with pulse shape analysis are considered (see Fig \ref{fig5}). No
indication for a peak at the Q-value $0\nu\beta\beta$-decay is
seen in none of the two data sets (the first 200 d of measurement
of each detector were suppressed, because of possible interference
with the cosmogenic $^{56}$Co). Data prior to PSD show a
background around the Q-value of ($0.19\pm0.1$)c/keV.kg.y similar
to that of IGEX (0.2c/keV.kg.y).

The average energy resolution at the Q$_{2\beta}$ value is
($4.23\pm0.14$)keV. The set of data analyzed with PSD correspond
to 35.5 kg.y of exposure and its background in the energy region
between 2000.2080 keV is ($0.06\pm0.01$)counts/(keV.kg.y).
Following the method proposed of PDG, the limit on the half-life
is \cite{X12}
\begin{center}
$T^{0\nu}_{1/2}\ge 1.9\times10^{25}$ y ~~~~~~~$90\%$ C.L.
\end{center}

After a new statistical analysis of the same set of data, some of
the authors of the H/M collaboration conclude \cite{klap} that
there exists evidence of a neutrinoless double beta peak. That
result, of scarcely $2\sigma$ statistical significance, has been
very widely contested \cite{feru,com}.





\begin{table}[ht]
\caption{Current best constraints on $\vert\langle m
\rangle\vert$. Results} \label{constraintsresults} \footnotesize
\begin{tabular}{ll}
\hline Experiment & $\vert\langle m \rangle\vert<$~~~(eV)\\
\hline
IGEX enrich. $^{76}$Ge&($0.33\sim1.35$)~(6.0kg)\\
H/M enrich. $^{76}$Ge&($0.35\sim1.05$)~(10.9kg)\\
MIBETA nat. $^{130}$Te&($0.85\sim2.1$)~(6.8kg)\\
\hline
\normalsize
\end{tabular}
\end{table}

\begin{table}[ht]
\caption{Expected best constraints on $\vert\langle m
\rangle\vert$.} \label{constraintsexpected} \footnotesize
\begin{tabular}{ll}
\hline Experiment & $\vert\langle m \rangle\vert<$~~~(eV)\\ \hline
NEMO3 enrich. $^{100}$Mo&0.1~(10kg)\\

CUORE nat. $^{130}$Te&0.05~(1Ton)\\

EXO enrich. $^{136}$Xe&0.02~(2Ton)\\

GEDEON-1 enrich. $^{76}$Ge& 0.04-0.10 (70kg)\\

GENIUS enrich. $^{76}$Ge&0.01~(1Ton)\\
 &0.001eV~~(10Ton)\\
MAJORANA enrich. $^{76}$Ge&0.01~(500kg)\\ MOON enrich.
$^{100}$Mo&0.03~(2Ton)\\ \hline \normalsize
\end{tabular}
\end{table}

In the case of other emitters, from the neutrinoless half-live
limits given in Table \ref{limits} one can derive Majorana
neutrino mass bound according to the nuclear model of his choice.
Table \ref{constraintsresults} shows the range of bounds derived
from the most sensitive experiment. Table \ref{models} shows the
neutrinoless half-lives theoretical predictions, in terms of
$\langle m_{\nu}\rangle$.

\section{Future Prospects: The Germanium Option}

It has been made clear that the neutrinoless double beta decays
are very rare phenomena which, if detected, will provide important
evidences of a New Physics beyond the Standard Model of Particle
Physics, and would have far-reaching consequences in Cosmology.
The experimental achievements accomplished during the last decade
in the field of ultra-low background detectors have lead to
sensitivities capable to search for such rare events. To increase
the chances of observing such rare events, however, large amounts
of detector mass are mandatory, but keeping the other experimental
parameters optimized.

The best Majorana neutrino effective mass bound obtained so far
have been obtained with Ge diodes: $\langle
m_{\nu}\rangle\leq$(0.3--1.3)eV (IGEX and H/M experiments).
Extension of these experiments are being considered with the
purpose of reaching the frontier of the 10 meV for the Majorana
neutrino mass bound.

The achievement of 0.2 c/(keV.kg.y) (at 2MeV), in the raw
background data of IGEX and H/M has been considered as a
stationary limit whose reduction requires further and deeper
investigation. Starting from the IGEX background achievements, the
main challenge of the future extended Ge experiments (besides
using large quantities of the emitter nucleus) is to substantially
improve the radiopurity of the detectors and components (both
intrinsic and induced) and to suppress, as best as possible, the
background originated from external sources. A final step in the
reduction could then be obtained through pulse shape analysis
(PSD) (a factor one third--one fourth) or by some new techniques.

The energy resolution, another important parameter in the search
for a neutrinoless signal --which should appear as a peak at the
two-electron summed energy at Q=2039 keV- is rather sharp in the
case of Ge-diodes ($3\sim4$ keV integrated along long period of
time), and so not too much progress can be expected. Needless to
say that special care must be put in maintain a long-term
stability in gain, resolution, noise, count rate, PSD signal, etc.

The success of the Germanium option in obtaining the most
stringent bound to the Majorana neutrino effective mass, the
mastering of the techniques in making Ge detectors, the ultra-low
background achievements (in their raw data spectrum), their
long-term stability, good energy resolution and reasonable nuclear
factor of merit, make the Germanium option --with conventional Ge
diodes- worth to be explored further, with improved experimental
parameters, in new generation experiments, with the objective of
reaching the 10-20 meV level for the Majorana neutrino effective
mass bound, where recent data from solar and atmospheric neutrino
experiments place the discovery potential of the neutrinoless
double beta decay.

Extensions of the respective IGEX and Heidelberg/Moscow
Ge-experiments to larger masses and hopefully lower backgrounds
are the Majorana project \cite{maj} (500 kg of enriched $^{76}$Ge)
and the GEDEON proposal \cite{X14} (70 kg of $^{76}$Ge in a first
step, followed by larger masses in subsequent steps). Both
proposals plan to use conventional Ge-diodes although with
modifications to improve substantially the background.

On the other hand, an extension of the H/M experiment is the
GENIUS project \cite{klap98}, which plan to use 1 ton of naked Ge
crystals embedded directly in liquid nitrogen in a cylinder
container. There exists also a proposal called GEM which would
locate 1 ton of naked HPGe diodes inside an spherical vessel
containing ultrapure liquid nitrogen in a water shield
(BOREXINO-CTF-like). Both experiments plan to reach a half-life
limit of T$_{1/2}^{0\nu} \ge 10^{28}$ y, which would lead to a
mass bound of 15 meV (in the case of using, for instance, one of
the QRPA nuclear matrix element, i.e., F$_{N}^{0\nu}=1.12\times
10^{-13}$ y$^{-1}$ for the Ge nuclear factor of merit \cite{X1}.

The Majorana project \cite{maj} will extend significantly the IGEX
mass (from 6.3 kg up to 500 kg of $^{76}$Ge), and starts from the
technical achievements of the IGEX detectors. It will use however
new, commercially developed segmented Ge detectors and new Pulse
Shape Discrimination (PSD) techniques developed after the
completion of IGEX. It will use isotopically enriched germanium
(86{\%} in $^{76}$Ge) as in IGEX, in an ensemble of 200 detectors
of about 2.5 kg each. Each detector is segmented into 12
electrically-independent volumes, each of which will be read-out
with the new designed PSD system. The two new ingredients (besides
the larger mass of germanium), i.e., the segmentation and the new
PSD, are supposed to reduce substantially the intrinsic background
(mainly cosmogenic) of the IGEX detectors, down to a 3.7{\%} only
(respectively 0.265 through PSD and 0.138 due to segmentation) of
the raw IGEX background in the Q$_{2\beta}$=2039 keV energy region
(B=0.2 c/(keV.kg.y)), down to 7.5$\times $10$^{-3}$ c/(keV.kg.y).
Capitalizing the fact that the IGEX background in this region was
accounted for as the decay of cosmogenic induced isotopes in the
Germanium ($^{68}$Ge, half-life of 271 days and $^{60}$Co,
half-life 5.7 years), the proponents of the Majorana experiment
anticipate that taking special care in the fabrication and
transport of the detectors and accounting for the decay of the
cosmogenic induced activities along the ten years of running time
of the experiment, a further suppression factor of about 20 could
be, presumably, obtained, and so the background expected would be
about B=5$\times $10$^{ - 4}$ c/(keV.kg.y) (corrected by the
efficiency of the PSD and segmentation cuts).

With the inputs Mt=5000 kg y, B=5$\times $10$^{-4}$ c/(keV.kg.y),
$\Gamma $=3.5 keV Majorana would obtain a half-life limit of
T$_{1/2}^{0\nu} \ge 4.2\times10^{28}$ y which provide bounds for
the Majorana neutrino mass ranging from $\langle
m_{\nu}\rangle\sim(20-70)$ meV according to the nuclear matrix
elements employed.

The GEDEON (GErmaniun DEtector in ONe cryostat) proposal
\cite{X14} is also an extension of the IGEX experiment, in various
steps including, in a first step, the use of natural abundance
germanium in a search for WIMPs. It also relies in the IGEX
technology, putting the emphasis mainly in the elimination of the
cosmogenic induced activity of the detectors, instead of in the
read-out. It will consist in sets of germanium crystals of 2.5 kg
each arranged in cells containing 28 Ge crystals (four planes of
seven Ge diodes per plane, in only one cryostat (cell-unit), made
in copper electroformed underground). A first cell, in natural
isotopic abundance germanium will test the performances of the
set-up and look for WIMPs. Then the inclusion of enriched
$^{76}$Ge detectors in cells of 70 kg would be implemented in
successive steps. The shielding will be made from Roman lead
(inner) and low activity lead (outer), in much the same way as in
IGEX. After that shield, neutron shieldings of polyethylene and
borated water as well as muon vetos would complete the set-up. The
experiment will be installed in the new Canfranc Underground
Laboratory. A quantitative study by MC simulations of the GEDEON
intrinsic background is now in progress; for instance, a
background of 0.002 c/keV/kg/day is reasonably expected in the low
energy region (Ref.\cite{X14}). Studies of the background in the 2
MeV region are in course.

GEDEON will use conventional (not-segmented) Ge detectors with
improved PSD. However, the main emphasis of this proposal will be
put in avoiding almost completely the activation of the crystals
and components. A significant background reduction is expected to
be obtained in GEDEON with respect to IGEX due to the growing of
the detector's crystals in the underground site itself. Also the
electroformed copper cryostats will be made underground, whereas
the multiple melting of archaeological lead and the machining of
bricks for shielding will be performed, in much the same way as in
IGEX, but underground. Clean rooms will be used for assembling and
mounting the detectors. The almost suppression of the cosmogenic
activation could lower the levels of raw background by a
conservative factor of 50, down to B=4$\times $10$^{-3}$
c/(keV.kg.y) from the starting "plateau" value of 0.2 c/(keV.kg.y)
of IGEX, that will be further reduced by PSD analysis by a factor
of 4, down to B$\sim 1\times10^{-3}$ c/(keV.kg.y). With this
value, and using the neutrinoless sensitivity expression for
germanium,

\begin{center}
S$_{1/2}^{0\nu} \sim4.94\times10^{24}$(Mt/B$\Gamma)^{1/2}$ years
\end{center}

\noindent one would obtain in the case of M=70 kg, $\Gamma$=3 keV,
t=5 years a half-life sensitivity bound of S$_{1/2}^{0\nu}
\sim1.7\times10^{27}$ y from where a bound of $\langle
m_{\nu}\rangle<40$ meV could be obtained by using the nuclear
matrix element calculation of \cite{X1}. Somehow higher values
would be obtained with other nuclear matrix elements (or
nuclear-factor-of-merit) from Table \ref{nuclearmerits}.

The Phase I of GEDEON could be considered as an intermediate step
($\langle m_{\nu}\rangle\ge(0.04\sim0.1)$ eV) conventional-Ge
experiment. The addition of more mass of emitters will provide
improved bounds, which could reach a sensitivity of
S$_{1/2}^{0\nu} \sim10^{28}$ y, with a set of five cells in 5
years. GEDEON offers as a bonus that the absence of cosmogenic
activation will permit to start the experiment with a very low
background without need to wait for the cooling of activations.

In conclusion, the Germanium detectors/emitters, which have
provided the best neutrino mass bounds, are a most valid option
for future, next generation double beta decay experiments.


\section{Conclusions and Outlook}

The standard two-neutrino decay mode has been directly observed in
several nuclei: $^{48}$Ca, $^{76}$Ge, $^{82}$Se, $^{96}$Zr,
$^{100}$Mo, $^{116}$Cd and $^{150}$Nd and others are under
investigation ($^{130}$Te, $^{136}$Xe) as summarized in Table
\ref{tablagorda}. Limits on neutrinoless half-lives have been
derived from that experiments (see the recent Reviews in Ref.
\cite{doblebeta}). Nuclear model calculations give still a large
spread of prediction, and that translates into a broad uncertainty
in the Majorana effective neutrino mass bound derived from the
results.

Data from the most sensitive experiments on neutrinoless double
beta decay lead to the limit $\langle m_\nu \rangle <0.3-1.3$ eV
for the effective neutrino mass, according to the nuclear model.
The Ge experiments provide the stringest bound to the neutrino
mass parameter and they seem to offer, for the next future, the
best prospectives to reach the lowest values of $\langle m_\nu
\rangle$. There exist obviously other emitters and techniques for
large next generation double beta decay experiments.

Currently running or nextcomming experiments (say, respectively,
NEMO 3 and CUORICINO) will explore effective neutrino masses down
to about 0.1---0.3 eV. To increase the sensitivity it is necessary
to go to larger source masses and reduce proportionally the
background, like the projects mentioned in Section 5. That would
bring the sensitivity to neutrino mass bounds down to a few
$10^{-2}$eV. Table \ref{constraintsexpected} give a simple look to
the expectations of some of these experiments, schematically
described in Table \ref{tablagorda}.

\section{ACKNOWLEDGEMENTS}

We thak Susana Cebri\'{a}n for her collaboration in the GEDEON
proposal and, in particular, for the MC estimation of the
background, and Frank Avignone for information and discussion on
the Majorana project. The financial support of CICYT (Spain) under
contract AEN99-1033 is acknowledged.

\onecolumn
\footnotesize
 \setlength\LTleft{0pt} \setlength\LTright{0pt}
\begin{longtable}[l]{l@{\extracolsep{\fill}}ll}
caca \kill
\caption{Synopsis of Double Beta Decay Searches}\label{tablagorda}\\
\hline
& \textbf{Method} & \textbf{Results}\\
 \hline \endfirsthead

\caption{(Continued)}\\
\hline
 & \textbf{Method} & \textbf{Results}\\
 \hline \endhead
\hline

& \textbf{$^{48}$Ca$\to^{48}$Ti Q=4276keV} & \\
HEP Inst. Beijing & Large scint. crystals of natural CaF$_{2}$(43
gr of Ca$^{48}$)&
T$_{1/2}^{0\nu}\ge1.14\times10^{22}$y (68\%)\\
 &
 &
T$_{1/2}^{0\nu\chi}>7.2\times10^{20}$y (90\%)\\
\\
UCI Hoover Damm & Helium TPC (atm. press.). Central powder source&
T$_{1/2}^{2\nu}=(4.3_{-1.1}^{+2.4}\pm1.4)\times10^{19}$y\\
260 mwe& CaCO$_{3}$ m=42.2 gr and m=10.3 g (73\% enriched) &
 \\
  &
efficiency $\sim 10\%$ &
 \\
\\ \hline
& \textbf{$^{76}$Ge$\to^{76}$Se Q=2039keV} & \\
Heidelberg/Moscow & Set of large (enriched) Ge detectors&
T$_{1/2}^{2\nu}=(1.55+0.01_{-0.15}^{+0.19})\times10^{21}$y\\
Gran Sasso 3330 mwe & 11.5 kg (10.9 kg fiducial) (2001)&
T$_{1/2}^{0\nu}\ge1.9\times10^{25}$y (90\%)\\
 &
 &
 T$_{1/2}^{0\nu\chi}\ge6.4\times10^{22}$y (90\%)\\
\\
IGEX & Set of large (enriched) Ge detectors&
T$_{1/2}^{2\nu}=(1.45\pm0.15)\times10^{21}$y \\
Canfranc 2450 mwe & 6.3 kg (6.0 kg fiducial)(1998)&
T$_{1/2}^{0\nu}\ge1.6\times10^{25}$y (90\%)\\
\\ \hline
& \textbf{$^{82}$Se$\to^{82}$Kr Q=2992keV} & \\
UCI Hoover Damm & Helium TPC (atm. press.). Central powder source&
T$_{1/2}^{2\nu}=(1.08_{-0.06}^{+0.26})\times10^{20}$y\\
260 mwe & (few grams) Efficiency $\sim10\%$)&
T$_{1/2}^{0\nu}\ge2.7\times10^{22}$y (68\%)\\
 &
 &
 T$_{1/2}^{0\nu\chi}\ge1.6\times10^{21}$y (68\%)\\
\\
NEMO 2 & Electron tracking device, Geiger cells. V=1 m$^{3}$, Eff
2\% & T$_{1/2}^{2\nu}=(0.83\pm0.09\pm0.06)\times10^{20}$y \\
Frejus 4800 mwe & External plastic calorim. Vertical central
source & T$_{1/2}^{0\nu}\ge9.5\times10^{21}$y (90\%)\\ (1998)&
enr. 97\% (156 g) and nat. (134 g) &
T$_{1/2}^{0\nu\chi}\ge2.4\times10^{21}$y (90\%)\\
\\
Heidelberg & Geochemical & T$_{1/2}=(1.30\pm0.05\times10^{20}$y\\
\\
Missouri & Geochemical &
T$_{1/2}=(1.0\pm0.4\times10^{20}$y\\
\\ \hline
& \textbf{$^{100}$Mo$\to^{100}$Ru Q=3351keV} & \\
INR Baksan  & Porportional chamber plus plasic scint.&
T$_{1/2}^{2\nu}=(3.3_{-0.1}^{+0.2})\times10^{18}$y\\
 660 mwe&
 Powdre/source interleaved. 46g (90\% enriched)&
  \\
\\
LBL Mt. Hol.  & Stack of semiconductors.&
T$_{1/2}^{2\nu}=(7.6_{-1.4}^{+2.2})\times10^{18}$y\\ UNM (Consil)
3300 mwe & 60g g(97\% enriched)&
T$_{1/2}^{0\nu}\ge2.2\times10^{22}$y (68\%)\\
\\
ELEGANTS V & Electron tracking det. Drift chambers, plastic scint.
&
T$_{1/2}^{2\nu}=(1.15_{-0.28}^{+0.30})\times10^{19}$y\\
Osaka (Kamioka) & NaI modules. Eff$\sim11\%(2\nu), 19\%(0\nu)$ &
T$_{1/2}^{0\nu}\ge6.5\times10^{22}$y (90\%)\\
2700 mwe & Enriched 104g (94.5\%) and nat. 171g &
T$_{1/2}^{0\nu\chi}\ge5.4\times10^{21}$y (68\%)\\
\\
UCI. Hoover Damm & Helium TPC (atm. press.) Central powdre source&
T$_{1/2}^{2\nu}=(1.16_{-0.08}^{+0.34})\times10^{19}$y \\
260 mwe &
 Enriched 8g (97.4\%) &
 T$_{1/2}^{0\nu}\ge2.6\times10^{21}$y (90\%)\\
 \\
 &
 Enriched 16.7g (97.4\%)&
 T$_{1/2}^{2\nu}=(6.82_{-0.53}^{+0.38}\pm0.68)\times10^{18}$y \\
 &
 Eff$\sim10\%(2\nu), 11\%(0\nu)$&
 T$_{1/2}^{0\nu}\ge1.23\times10^{21}$y (90\%)\\
 &
 &
 T$_{1/2}^{0\nu\chi}\ge3.31\times10^{20}$y\\
 \\
 NEMO 2 Frejus&
 Electron tracking detector&
 T$_{1/2}^{2\nu}=(0.95\pm0.04\pm0.09)\times10^{19}$y \\
 4800 mwe&
 Enriched 172g (98.4\%). Nat. 163g&
 T$_{1/2}^{0\nu}\ge6.4\times10^{21}$y (90\%)\\
 &
 Eff. $\sim$ few percent&
 T$_{1/2}^{0\nu\chi}\ge5.0\times10^{20}$y (90\%)\\
 \\ \hline
& \textbf{$^{96}$Zr$\to^{96}$Mo Q=3351keV} & \\
Kawashima et al.& Geochemical. $1,7\times10^{9}$y old Zircon&
T$_{1/2}=(3.9\pm0.9)\times10^{19}$y\\
\\
NEMO 2 Frejus&
 Electron tracking detector. Central vertical source&
 T$_{1/2}^{2\nu}=(2.1_{-0.4}^{+0.8}\pm0.2(syst))\times10^{19}$y \\
 4800 mwe&
 Enriched 20.5g (57.3\%) ZrO$_{2}$. Nat 18.3g&
 T$_{1/2}^{0\nu}\ge1.3\times10^{21}$y (90\%)\\
 &
 t=10357 h&
 T$_{1/2}^{0\nu\chi}\ge3.5\times10^{20}$y (90\%)\\
 \\ \hline
& \textbf{$^{116}$Cd$\to^{116}$Sn Q=2804keV} & \\ ELEGANTS V &
Electron tracking det. Drift chambers, plastic scint.&
T$_{1/2}^{2\nu}=(2.6_{-0.5}^{+0.9}\pm0.35)\times10^{19}$y\\ Osaka
(Kamioka)& Enriched 91.1g (90.7\%). Nat 88.5g, Eff. 8\%&
T$_{1/2}^{0\nu}\ge5.44\times10^{21}$y (90\%)\\
\\
NEMO 2 Frejus&
 Electron tracking detector. Central vertical source&
 T$_{1/2}^{2\nu}=(3.75\pm0.35\pm0.21)\times10^{19}$y \\
 4800 mwe&
 Enriched 152g (93.2$\%$). Nat 143g&
 T$_{1/2}^{0\nu}\ge5.0\times10^{21}$y (90\%)\\
 &
 Efficiency $\sim1.7\%$ &
 T$_{1/2}^{0\nu\chi}\ge1.2\times10^{21}$y (90\%)\\
\\
 INR Kiev &
 Cadmium Tungstate scint. crystal &
 T$_{1/2}^{2\nu}=(2.7_{-0.5-0.6}^{+0.5+0.9})\times10^{19}$y \\
 (Solotvina) 1000 mwe  &
  CdWO$_{4}$ enriched 83\% and natural &
  T$_{1/2}^{0\nu}\ge 7\times10^{22}$y (90\%) \\
   &
  Efficiency 83.5\% ($0\nu$) &
   T$_{1/2}^{0\nu\chi}\ge1.2\times10^{21}$y (90\%)\\
\\ \hline
& \textbf{$^{128,130}$Te$\to^{128,130}$Xe Q=867/2529keV} & \\
Washington. St Louis & Geochemical. 4g. $2\times10^{9}$y old
Tellurium ore &
T$_{1/2}^{130}$/T$_{1/2}^{128}=(3.52\pm0.11)\times10^{-4}$\\
  & & T$_{1/2}^{130}=(2.7\pm0.1)\times10^{21}$y\\
   & & T$_{1/2}^{128}=(7.7\pm0.4)\times10^{24}$y\\
    & & T$_{1/2}^{130}$ controversial. Instead\\
       & & $(7-27)\times10^{20}$y\\
    & & T$_{1/2}^{0\nu}(128)>6.9\times10^{24}$y\\
    & & T$_{1/2}^{0\nu}(130)>2.54\times10^{21}$y\\
\\
Missouri & Geochemical & T$_{1/2}^{130}=(0.75\pm0.3)\times10^{21}$y\\
& & T$_{1/2}^{128}=(1.4\pm0.4)\times10^{24}$y\\
 & 1991. 5g $3\times10^{19}$y old Tellurium ore&
 T$_{1/2}^{130}$/T$_{1/2}^{128}=(4.2\pm0.8)\times10^{-4}$\\
\\
 Heidelberg & Geochemical & T$_{1/2}^{130}=(1.5-2.8)\times10^{21}$y \\
  & &  T$_{1/2}^{128}>5\times10^{24}$y\\
\\
Takaoka et al. & Geochemical & T$_{1/2}^{130}=(7.9\pm1.0)\times10^{20}$y\\
 & & T$_{1/2}^{128}=(2.2\pm0.3)\times10^{24}$y\\
 & & from T$_{1/2}^{130}$/T$_{1/2}^{128}=(3.5\pm0.11)\times10^{-4}$\\
 \\
 recommended values:& T$_{1/2}^{130}=8\times10^{20}$y ;
  T$_{1/2}^{128}=(2\times10^{24}$y\\
 \\ \hline
& \textbf{$^{130}$Te$\to^{130}$Xe Q=2529keV} & \\ Milan  &
Cryogenic exp. Bolometers of TeO$_{2}$ & \\ MIBETA &Natural Te
(34\%). NTD sensors. Eff $\sim100\%$ & \\ (Gran Sasso)
 & 73 g at 15 mK & T$_{1/2}^{0\nu}\ge2.5\times10^{21}$y (90\%)\\
 & 334 g at 10 mK & T$_{1/2}^{0\nu}\ge2.8\times10^{22}$y (90\%)\\
 & 4x330 g at 10 mK & T$_{1/2}^{0\nu}\ge2.4\times10^{22}$y (90\%)\\
 & 20x340 g   & T$_{1/2}^{0\nu}\ge2.1\times10^{23}$y (90\%)\\
  & & T$_{1/2}^{0\nu}(^{128}Te\ge1.7\times10^{22}$y (90\%)\\
\\
& \textbf{$^{136}$Xe$\to^{136}$Ba Q=2467keV} & \\
Caltech/Neuchatel/PSI &
 Xe TPC (enr. 62.5\%) at 5 bar. Fiducial 180 l. 24.4 mol &
 T$_{1/2}^{2\nu}\ge5.5\times10^{20}$y (90\%)  \\
Gothard 3000mwe &
 (3.3 kg of $^{136}$Xe) Global eff. 22\%, $\Gamma$(2480 keV)$\sim6.6$\%&
 T$_{1/2}^{0\nu}\ge4.4\times10^{23}$y (90\%)\\
 &Calorimeter and tracking detector at the same time &
 T$_{1/2}^{0\nu\chi}\ge7.2\times10^{21}$y (90\%)\\
Rome 2 (Gran Sasso) & Xenon chamber, Enriched source (2001) &
T$_{1/2}^{0\nu\chi}\ge7\times10^{23}$y \\
\\ \hline
& \textbf{$^{150}$Nd$\to^{150}$Sm Q=3368keV} & \\
UCI. Hoover Damm & Helium TPC (atm. press.) Central powder source&
T$_{1/2}^{2\nu}=(6.75_{-0.42}^{+0.37}\pm0.68)\times10^{18}$y \\
260 mwe &
 Source 15.5 g of Nd$_{2}$O$_{3}$ enr.(91\%). 1200 Gauss &
 T$_{1/2}^{0\nu}\ge1.22\times10^{21}$y (90\%)\\
  & Efficiency $\sim11\% (0\nu, 2\nu)$ &
  T$_{1/2}^{0\nu\chi}\ge2.82\times10^{20}$y (90\%)\\
 \\
ITEP &
  CH$_{4}$ TPC (atm. press.). 800 Gauss. Prototype 0.3m$^{3}$ &
   T$_{1/2}^{2\nu}=(1.88_{-0.37}^{+0.69}\pm0.19)\times10^{19}$y\\
at sea level & 40 g of $^{150}$Nd (92\% enr.) and nat. 2.5 g of
$^{150}$Nd&
   T$_{1/2}^{0\nu}\ge2.1\times10^{20}$y (90\%)\\
and in Baksan & Efficiency $\sim3\%$ &
T$_{1/2}^{0\nu\chi}\ge1.7\times10^{20}$y (90\%)\\
\\ \hline
& \textbf{$^{238}$U$\to^{238}$Pu Q=1437keV} & \\
Chicago/ & Radiochemical experiment. Milking to get 5.5 MeV
$\alpha$'s &
T$_{1/2}=(2.0\pm0.6)\times10^{21}$y \\
Santa Fe/ & from $^{238}$Pu (90y). Source 8.47g uranyl nitrate &
 T$_{1/2}^{0\nu}>0.84\times10^{21}$y\\
Los Alamos  &Ingrowth time: 33 y &
 T$_{1/2}^{0\nu\chi}>0.8\times10^{21}$y\\
\\ \hline
& \textbf{$^{244}$Pu$\to^{244}$Cm Q=1352keV} & \\
Lawrence &Radiochemical experiment. Milking to get
5.8MeV$\alpha$'s  &
 T$_{1/2}\ge1.1\times10^{18}$y  (95\%)\\
Livermore & from $^{244}$Cm (18y). Source 1.5g &
 T$_{1/2}^{0\nu}>1.1\times10^{18}$y\\
National Lab.  &Ingrowth time: 1.08 y &
 T$_{1/2}^{0\nu\chi}>1.1\times10^{18}$y\\
\\
 \hline
\end{longtable}
\normalsize \twocolumn

\end{document}